\documentclass[pre,showpacs]{revtex4}
\usepackage{amssymb}
\usepackage[dvips]{graphicx}
\usepackage{amsmath}
\usepackage{graphicx}
\usepackage{makeidx}
\usepackage{subfigure}
\begin{document}
\title{A coupled map lattice model for spontaneous pore formation in anodic oxidation}
\author{Hidetsugu Sakaguchi and Jie Zhao}
\affiliation{Department of Applied Science for Electronics and Materials,
Interdisciplinary Graduate School of Engineering Sciences, Kyushu
University, Kasuga, Fukuoka 816-8580, Japan}

\begin{abstract}
We  construct a coupled map lattice model for the spontaneous pore formation in anodic oxidation in two dimensions and perform numerical simulations, after we explain steady flat solutions, their linear stability and a single pore solution for a model of Parkhutik and Shershulsky. 
\end{abstract}
\maketitle
\section{Introduction}
Nanoscale pores are spontaneously created in alumina film, when anodizing aluminum is oxidized to alumina (Al$_2$O$_3$) in various acid solutions~\cite{rf:1}. A schematic figure of the pore structure is drawn in Fig.~1(a). 
The anodic aluminum is used as aluminum pots and aluminum window sash in daily life. The alumina film works as a protection film against further corrosion. The configuration of pores is usually irregular, but regular arrays of pores were found in controlled experiments~\cite{rf:2,rf:3}.  
Similar porous structures were reported in anodic titanium~\cite{rf:4}. The mechanism of the pore formation in anodic oxidation has been long studied, but is not still completely understood~\cite{rf:5}. Recently, the porous anodic films were studied from a view point of the instability of growing interfaces~\cite{rf:6,rf:7,rf:8,rf:9,rf:10}. Unstable growing interfaces have been intensively studied in the problems of dendrites and viscous fingering~\cite{rf:11}.  The feature in the problem of the anodic oxidation is that there are two interfaces between Al and Al$_2$O$_3$ and Al$_2$O$_3$ and acid solution (electrolyte), and their interaction is important. The linear stability analysis for flat interfaces was performed by Thamida and Chang~\cite{rf:7}, but the model does not provide a physically justified short-wave cutoff. An elastic effect by the volume change at the oxidation can play an important role for the formation of the regular hexagonal arrays of pores~\cite{rf:3}. Singh proposed a more plausible model including the elastic effect,  performed the linear stability analysis, and derived a weakly nonlinear equation similar to the Kuramoto-Sivashinsky equation~\cite{rf:8,rf:9}.  

However, the well-developed pores were not well treated because of the difficulty of the moving boundary conditions.  
The Stefan problem with the moving boundary conditions has been studied typically in the  problem of dendritic crystal growth. 
There are several models to treat well-developed complicate interface patterns.  Witten and Sander proposed a DLA (diffusion-limited aggregation) model for fractal growth patterns~\cite{rf:12}.    
The phase field model was proposed to perform numerical simulations for dendritic patterns~\cite{rf:13}. We proposed a coupled map lattice model as a simpler simulation model of dendrites and extended it the oscillatory electrodeposition and the pore formation in activated carbon~\cite{rf:14,rf:15,rf:16}. In this paper, we study a simple model for the spontaneous pore formation in anodic oxidation which was first proposed by Parkhutik and Shershulsky, and perform numerical simulations using a coupled map lattice model. The model does not include the elastic effect and cannot reproduce a regular array of pores. However, the model can simulate a strongly nonlinear evolution toward the well-developed pores and  competitive dynamics among the neighboring developed pores.  

\section{Simple model for anodic oxidation and its linear stability analysis}
The anodic oxidation of aluminum creates alumina film on the surface of aluminum.
Flat alumina films are stable in alkali solutions but they are unstable to form many pores in acid solutions such as sulfuric acid solutions. It is known that the thickness $d$ of the alumina film is proportional to the voltage $V$ of the anode and the average interval between neighboring pores increases with $V$. The detailed processes of chemical reactions are not well known but the chemical reaction:
\begin{equation}
2{\rm Al}+3{\rm OH}^{-}\rightarrow {\rm Al}_2{\rm O}_3+3{\rm H}^{+}+6{\rm e}^{-}
\end{equation}
occurs at the interface between Al and  Al$_2$O$_3$, and the alumina dissolves chemically at the interface between Al$_2$O$_3$ and the electrolyte. 
As a result of the chemical reactions, the interface patterns as shown in Fig.~1(a) move downwards, i.,e., in the $-z$-direction. 
The chemical reaction rates depend on the voltage at the interface. The electric current is induced by the chemical reactions. We study a simple dynamical model for the metal-oxide and oxide-electrolyte interfaces. 
The positions of the two interfaces are respectively expressed as $\zeta_1({\bf x},t)$ and $\zeta_2({\bf x},t)$, whose schematic figure is drawn in Fig.~1(b).  The electric potential is denoted as $\phi$. The voltage is assumed to be $\phi=V$ at the first interface $\zeta_1$. The electric potential satisfies
\begin{equation}
\nabla^2\phi=0.
\end{equation}
The growing velocity of the first interface is written as  
\begin{equation}
\frac{d\zeta_1}{dt}=-aE_e,
\end{equation}
where $E_e$ is the electric field strength, $a>0$  is a Faradaic coefficient multiplied by the conductivity of the oxide, and $\partial_n$ implies the normal differentiation. Parkhutik and Shershulsky proposed a model for the velocity of the second interface at $\zeta_2$ as 
\begin{equation}
\frac{d\zeta_2}{dt}=-\alpha_0\exp(k_dE_e)+\beta_0\exp(k_oE_e),
\end{equation}
where $\alpha_0,\beta_0,k_d$ and $k_o$ are some coefficients for dissolution and oxidation. 
Here, we propose a simpler model, assuming that the electric field is weak near $\zeta_2$:
\begin{equation}
\frac{d\zeta_2}{dt}=-b_0-b_1E_e+b_2E_e^2+D\kappa,
\end{equation}
where $b_0,b_1$and $b_2$ are expansion coefficients of the interface velocity with $E_e$, $\kappa=(\partial^2/\partial x^2+\partial^2/\partial y^2)\xi_2$ denotes the curvature of the interface, and $D$ is a coefficient proportional to the surface tension. The interface tends to become flat by the effect of the surface tension. The electric potential is assumed to $\phi=0$ at $z=\xi_2$. 

\begin{figure}[tbp]
\begin{center}
\includegraphics[height=3.5cm]{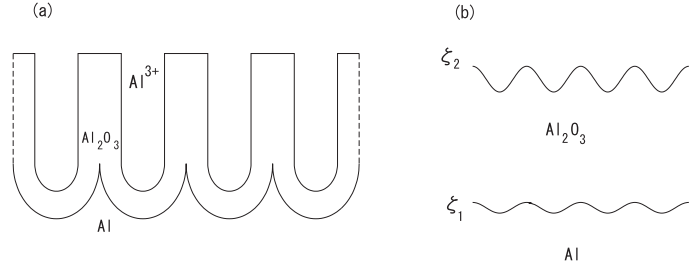}
\end{center}
\caption{(a) Schematic figure of porous alumina. (b) Two interfaces of aluminum-alumina and alumina-electrolyte.}
\label{f1}
\end{figure}
If the flat interfaces are stable, the electric potential satisfies $\phi=V(\zeta_2-z)/(\zeta_2-\zeta_1)$, and $\partial_n\phi=-\partial \phi/\partial z=V/d$, where $d=\zeta_2-\zeta_1$ is the interval between the two interfaces. The velocity $d\zeta_1/dt=-v_0=-aV/d$ and the velocity $d\zeta_2/dt=-v_0=-b_0-b_1(V/d)+b_2(V/d)^2$. The steady electric field $E=V/d$ is determined from $d\zeta_1/dt=d\zeta_2/dt$ as
\begin{equation}
\frac{V}{d}=\frac{b_1-a+\sqrt{(b_1-a)^2+4b_0b_2}}{2b_2}.
\end{equation}
This relation shows that the thickness $d$ is proportional to the voltage $V$. 
If $b_2=0$, $V/d=b_0/(a-b_1)$, and $b_0>0$ and $a>b_1$ are necessary for 
stable growth. If $b_0=0$, $V/d=(b_1-a)/b_2$, and $b_2>0$ and $b_1>a$ are necessary for stable growth.

The stability of the flat interfaces can be investigated, assuming the perturbation of the form: 
\begin{eqnarray}
\delta\phi&=&V_0(d-z-v_0t)/d+\delta\phi_q^1\cos(qx)e^{-qz+\lambda_qt}+\delta\phi_q^2\cos(qx)e^{qz+\lambda_q t},\nonumber\\
\zeta_1&=&-v_0t+a_q\cos(qx)e^{\lambda_q t},\;\;\zeta_2=-v_0t+d+b_q\cos(qx)e^{\lambda_qt}.
\end{eqnarray}
\begin{figure}[tbp]
\begin{center}
\includegraphics[height=3.5cm]{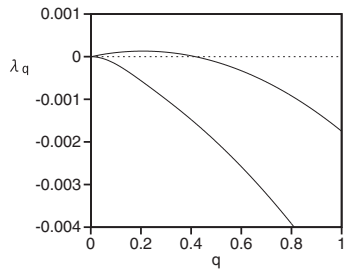}
\end{center}
\caption{Eigenvalue $\lambda_q$ at $b_1=0.035$ and 0.02 for $a=0.015,b_0=0,,b_2=0.04$ and $D=0.003$.}
\label{f2}
\end{figure}
From the boundary conditions at the two interfaces, 
\begin{eqnarray}
& &-(V/d)a_q+\delta\phi_q^1+\delta\phi_q^2=0,\nonumber\\
& &-(V/d)b_q+\delta\phi_q^1e^{-qd}+\delta\phi_q^2e^{qd}=0,\nonumber\\
& &\lambda_qa_q+aq(\delta\phi_q^1-\delta\phi_q^2)=0,\nonumber\\
& &(\lambda_q+Dq^2)b_q+\{b_1+2b_2(V/d)\}(e^{-qd}\delta\phi_q^1-e^{qd}\delta\phi_q^2)=0.
\end{eqnarray}
The eigenvalue $\lambda_q$ satisfies
\begin{eqnarray}
& &(e^{qd}-e^{-qd})\lambda_q^2+[\{Dq^2+aq(V/d)-b_1q(V/d)-2b_2q(V/d)^2\}e^{qd}-\{Dq^2-aq(V/d)+b_1q(V/d)+2b_2q(V/d)^2\}e^{-qd}]\lambda_q\nonumber\\
& &+aq(V/d)[\{Dq^2-b_1q(V/d)-2b_2q(V/d)^2\}e^{qd}+\{Dq^2+b_1q(V/d)+2b_2(V/d)^2\}e^{-qd}]=0.
\end{eqnarray}
Figure 2(a) shows $\lambda_q$ at $b_1=0.02$ and 0.035 for $a=0.015,b_0=0,b_2=0.04$ and $D=0.003$. The flat interfaces is stable at $b_1=0.035$, but they are unstable at $b_1=0.02$. This instability is analogous to the Mullins-Sekerka instability in crystal growth, however, the interaction between the two interfaces is important in this problem. 

\section{A single pore solution}
When the flat interfaces become unstable, many pores are spontaneously created. 
A single pore solution can be explicitly constructed for $b_0=b_2=0$ in two dimensions as a generalization of the Saffman-Taylor solution in a channel for the viscous fingering~\cite{rf:17,rf:18}.  
Single pore solutions for more complicated conditions such as Eq.~(4) were numerically solved by Thamida and Chang~\cite{rf:7}. The potential satisfies $(\partial_{xx}+\partial_{zz})\phi=0$, and the boundary  conditions are $a\partial_n\phi=U\cos\theta_1$ at the first interface and $b_1\partial_n\phi=U\cos\theta_2$, where $U$ is the uniform propagation velocity in the $-z$ direction, and $\theta_{1,2}$ are the angles between the normal directions of the two interfaces and the $z$ direction.  
For a spatially-periodic array of pores with wavelength $2x_0$, $2x_0$ corresponds to the channel width in the Saffman-Taylor solution. 
Another variable $\psi$ is introduced, where $\phi+i\psi$ is a holomorphic function in the complex plane $-z+ix$. From the boundary conditions $a\partial_n\phi=U\cos\theta_1$, $b_1\partial_n\phi=U\cos\theta_2$, $\partial \psi/\partial x=U/a$ at the first interface, and $\partial \psi/\partial x=U/b_1$ at the second interface, which yields 
\begin{figure}[tbp]
\begin{center}
\includegraphics[height=4.cm]{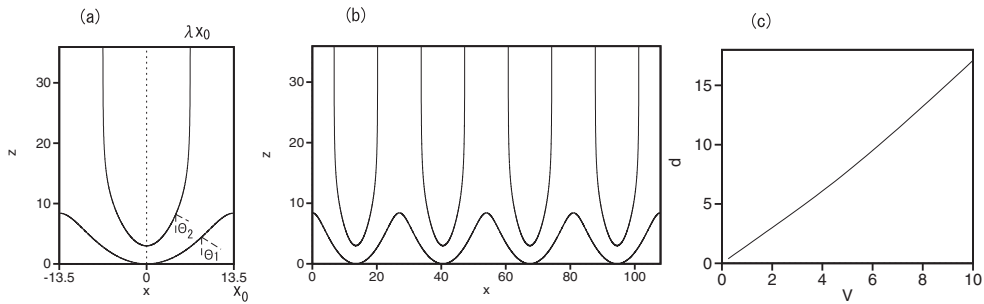}
\end{center}
\caption{(a) A single pore solution at $\lambda=0.5,x_0=13.5,V=2$ and $b_1/a=0.5$. (b) An array of pore structures. (c) Thickness $d$ as a function of $V$ for $\lambda=0.5,x_0=13.5$ and $b_1/a=0.5$ }
\label{f3}
\end{figure}
$\psi=Ux/a$ at the first interface, and $\psi=Ux/b_1$ at the second interface.  Furthermore, $\phi=V$ at the first interface, $\phi=0$ at the second interface, and $\psi=\pm V_0$ is assumed at the boundaries $x=\pm x_0$. 
The width of the pore is assumed to be $2\lambda x_0$.  
The schematic figure is shown in Fig.~3(a). 
The inverse function $x(\phi,\psi)$ can be expanded with the Fourier series as
\begin{equation}
\frac{x}{x_0}=\frac{\psi}{V_0}+\sum\{A_n\sin(n\pi\psi/V_0)e^{-n\pi\phi/V_0}+B_n\sin(n\pi\psi/V_0)e^{n\pi\phi/V_0}\}.
\end{equation}
At the first interface, $\psi=Ux/a$ and $\phi=V$, which yields
\begin{equation}
\frac{a\psi}{x_0U}=\frac{\psi}{V_0}+\sum \{A_ne^{-n\pi V/V_0}+B_ne^{n\pi V/V_0}
\}\sin(n\pi \psi/V_0).
\end{equation}
At the intersections of the first interface and the boundaries $x=x_0$, $\psi$ takes $V_0$ owing to the continuity, and therefore $a/U=x_0/V_0$, and 
\begin{equation}
B_n=-e^{2n\pi V/V_0}A_n.
\end{equation}
At $x=\lambda x_0$ and $z=\infty$, $\psi$ approaches $V_0$. It leads to $U\lambda x_0/b_1=V_0$ and  $\lambda=b_1/a$.
At the second interface, $\psi=Ux/b_1$ and $\phi=0$, which yields
\begin{equation}
\frac{(\lambda-1)\psi}{x_0U}=\sum(A_n+B_n)\sin(n\pi\psi/V_0).
\end{equation}
From the Fourier expansion (13) and the relation (12), 
\begin{equation}
A_n=\frac{2(-1)^n(1-\lambda)}{(n\pi)\{1-\exp(-2n\pi V/V_0)\}}.
\end{equation}
Using the Cauchy-Riemann relation: $\partial x/\partial \psi=-\partial z/\partial \phi$, the first interface satisfying $\phi=V$ is expressed as
\begin{equation}
\frac{z}{x_0}=-\frac{V}{V_0}+\sum(2A_n)e^{-n\pi V/V_0}\cos(n\pi x/x_0)+C,
\end{equation}
where an integral constant $C$ is determined as $C=V/V_0+\sum(-2A_n)e^{-n\pi V/V_0}$ by assuming $z=0$ at $x=0$. The second interface satisfying $\phi=0$ is expressed as
\begin{equation}
\frac{z}{x_0}=\sum\{(A_n-B_n)\cos\{n\pi x/(\lambda x_0)\}+C.
\end{equation} 
Figure 3(a) is an example of pore solution for $\lambda=0.5,x_0=13.5,V=2$ and $b_1/a=0.5$.  Figure 3(b) shows an array of pore structures obtained by repeating the single pore solution by wavelength $2x_0$. The mirror-symmetric spatially-periodic solution is also a special solution satisfying the boundary conditions.  
The pore solution by Thamida nd Chang seems to have a sharp cusp at the boundaries \cite{rf:7},  however, our solution has no such singularity.  
Figure 3(c) shows a relation of the thickness $d$ of the two interfaces at $x=0$ as a function of $V$. A monotonic increase of $d$ is observed. 
When the thickness $d$ is large, the first interface becomes flat and the second interface take a form similar to the Saffman-Taylor solution.

\begin{figure}[tbp]
\begin{center}
\includegraphics[height=4.5cm]{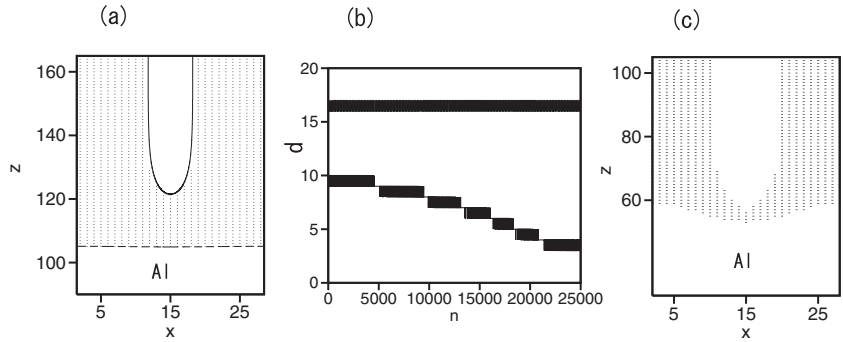}
\end{center}
\caption{(a) Pore structure at $a=0.042,b_0=b_2=g=0, b_1=0.01$ and $V=5$ starting from $d(0)=16$. The system size is $35\times 400$. The shaded region represents oxide. (b) Time evolution of the thickness $d(t)$. (c) Pore structure at $a=0.042,b_0=b_2=g=0, b_1=0.01$ and $V=5$ starting from $d(0)=10$
}
\label{f4}
\end{figure}
\begin{figure}[tbp]
\begin{center}
\includegraphics[height=4.5cm]{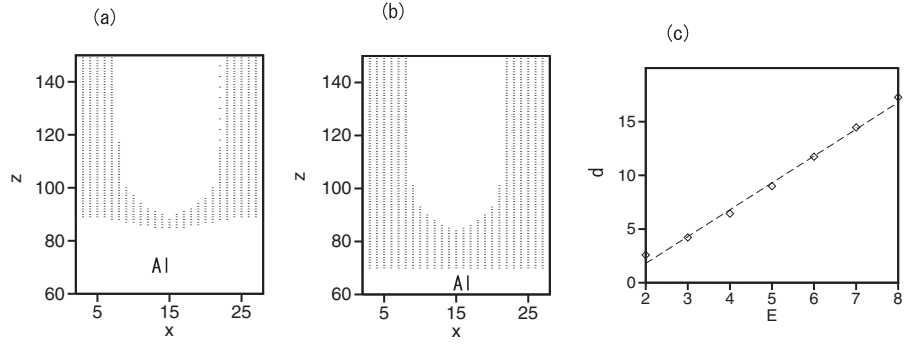}
\end{center}
\caption{(a) (b) Pore structures at (a) $V=3$ and (b) $7$ for $a=0.02,b_0=g=0,b_1=0.01$ and $b_2=0.005$.  5(c) Relation of the thickness $d$ and $V$. The dashed line is a linear fitting line. }
\label{f4}
\end{figure}
\begin{figure}[tbp]
\begin{center}
\includegraphics[height=4.cm]{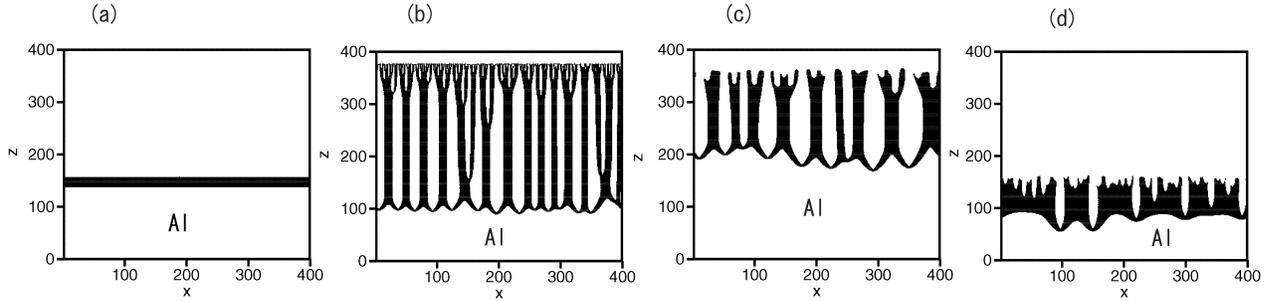}
\end{center}
\caption{(a) Flat pattern at $b_0=0,a=0.015,b_1=0.035,b_2=0.04,g=0.0003$ and $V=12$. (b) Pore structure at $a=0.02,b_0=0,b_1=0.01,b_2=0.005,g=0$ and $V=2$.
(c) Pore structure at $a=0.02,b_0=0,b_1=0.01,b_2=0.005,g=0.005$ and $V=2$.
(d) Pore structure at $a=0.045,b_0=0.0002,b_1=0.01,b_2=0,g=0$ and $V=3$.}
\label{f6}
\end{figure}
\begin{figure}[tbp]
\begin{center}
\includegraphics[height=4.cm]{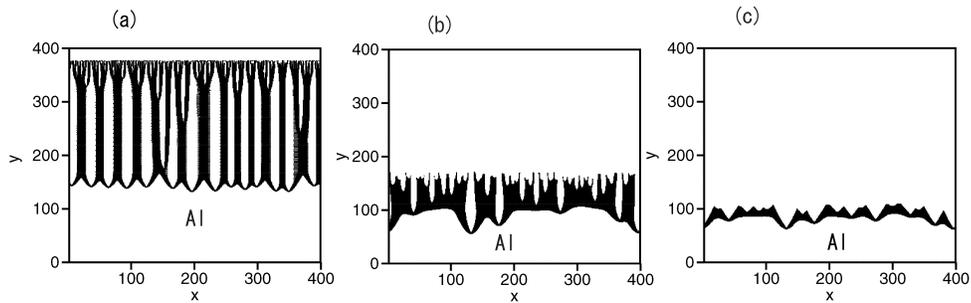}
\end{center}
\caption{(a)Pore structure at $a=0.02,\alpha_0=0.01,k_d=0,\beta_0=0.01,k_o=-1$ and $V=2$.
(b) Pore structure at $a=0.045,\alpha_0=0.05,k_d=0.2,\beta_0=0.0498,k_o=0$ and $V=3$.
(c) Pore structure at $a=0.045,\alpha_0=0.01,k_d=1.2,\beta_0=0.005,k_o=1.1$ and $V=3$ Pore structure at $a=0.02,\alpha_0=0.01,k_d=0,\beta_0=0.01,k_o=-1$ and $V=2$.}
\label{f7}
\end{figure}
\section{Numerical simulations with a coupled map lattice model}
We propose a coupled map lattice model on a square lattice corresponding to the model expressed by Eqs.~(2), (3) and (5) to investigate the nonlinear time evolution of the anodic oxidation. The time and the space are discretized in the coupled map lattice model.  
The discrete Laplace equation is written as 
\begin{equation}
\phi_n(i+1,j)+\phi_n(i-1,j)+\phi_n(i,j+1)+\phi_n(i,j-1)-4\phi_n(i,j)=0.
\end{equation}
We have solved the discrete Laplace equation by iteration of a hypothetical discrete diffusion equation.
\begin{equation}
\phi_n^{m+1}(i,j)=\phi_n^{m}(i,j)+D_{\phi}\{\phi_n^{m}(i+1,j)+\phi_n^n(i-1,j)+\phi_n^m(i,j+1)+\phi_n^m(i,j-1)-4\phi_n^m(i,j)\},
\end{equation}
with $D_{\phi}=0.2$, and the iteration number of $m$ is set to be 400 in most numerical simulations.  
The potential $\phi$ is set to be $\phi=V$ in the metal region and $\phi=0$ in the electrolyte region. We introduce two kinds of order parameters $p$ and $q$. The order parameters are $p=0$ and $q=1$ in the metal region, $p=1$ and $q=1$ in the oxide region, and $p=1$ and $q=0$ in the electrolyte region.  
The order parameter changes in time only at the interface sites.
In the interface sites between metal and oxide, $p$ takes a value between 0 and 1, and 
$p$ increases as 
\begin{equation}
p_{n+1}(i,j)=p_n(i,j)+a|\partial_n\phi_n(i,j)|,
\end{equation}
where $|\partial_n\phi|$ is numerically evaluated with 
\[|\partial_n\phi|=\sqrt{\{(\phi(i+1,i)-\phi(i,j))^2+(\phi(i-1,j)-\phi(i,j))^2+(\phi(i,j+1)-\phi(i,j))^2+(\phi(i,j-1)-\phi(i,j))^2\}/2},\]
because $(\nabla\phi)^2=(\partial_n\phi)^2+(\partial_s\phi)^2$ and the differential  $\partial_s\phi$ along the interface is zero. 
If $p_{n+1}(i,j)$ goes over 1, the interface site changes into a metal site. 
In the interface sites between oxide and electrolyte, $q$ takes a value between 0 and 1, and $q$ decreases as
\begin{equation}
q_{n+1}(i,j)=q_n(i,j)-\{b_0+b_1E_{en}(i,j)-b_2E_{en}(i,j)^2+g(N_{oc}(i,j)-3)\},
\end{equation}
where $E_{en}=|\partial_n\phi|$ is approximated at $\phi_n(i,j)$ because $\phi_n(i,j)$ is 0 in the electrolyte region, $N_{oc}(i,j)$ is the number of the electrolyte sites which locate at the nearest and the next-nearest neighbor sites of $(i,j)$, and $g$ is a coefficient corresponding to $D$ in Eq.~(5). The term $N_{oc}-3$ is used to represent the surface tension effect, because $N_{oc}$ is 3 for flat interfaces and the surface tension effect is zero for the flat interfaces. This term was used in the coupled map lattice model for dendritic growth including the surface tension effect~\cite{rf:12}.  
If $q_n(i,j)$ goes below 0, the interface site changes into an electrolyte site. 
Figure 4(a) displays the oxide region obtained by the coupled map lattice model with system size $27\times 400$ for $a=0.042,b_0=b_2=g=0, b_1=0.01$ and $V=5$. The solid and dashed lines are the analytic single pore solution obtained in the previous section. Fairly good agreement is seen. Figure 4(b) shows the time evolutions of the thickness $d$ between the first and the second interfaces, when the initial value of $d(0)$ is changed. When $d(0)=16$, the thickness $d(t)$ is almost constant in time, however, $d(t)$ decreases in time for $d(0)=10$. Figure 4(a) shows a pore structure for $d(0)=16$. Figure 4(c) displays the oxide region for $d(0)=10$ at $n=25000$. The steady pore solution by Eqs.~(15) and (16) is not always realized as a stable solution when $b_0=b_2=0$.  

If $b_0=0$ and $b_2>0$, a unique stable pore solution is obtained, because too small $d$ makes the velocity of the second interface slower and $d$ becomes larger. Figures 5(a) and (b) display the oxide region obtained by the coupled map lattice model of $27\times 400$ at $V=3$ and $7$ for $a=0.02,b_0=g=0,b_1=0.01$ and $b_2=0.005$.  Figure 5(c) displays a relation of the stationary value of the thickness $d$ as a function of $V$ obtained by the coupled map lattice model. A monotonously increasing relation is obtained. The single pore structure is similar to the Saffman-Taylor solution, when $d$ is relatively large. 
   
In numerical simulations shown in Fig.~4 and 5, a small dent is set in the second interface around $x=15$ as an initial condition. In Fig.~6, we show some numerical results in a larger system of $400\times 400$. The initial conditions are flat interfaces with small random perturbation.  Figure 6(a) shows stable flat interfaces for $b_0=0,a=0.015,b_1=0.035,b_2=0.04,g=0.0003$ and $V=12$. The flat interfaces are stable for the parameter values as shown in \S 2. Figure 6(b) shows a pore pattern for $a=0.02,b_0=0,b_1=0.01,b_2=0.005,g=0$ and $V=2$ at $n=60000$. Fine pore structures are created initially, because $g=0$, and it remains on the top region.  A kind of competition occurs among pores, and some pores grow and the others stop to grow. After all, a rather regular array of pores appears below $z=250$.  
Qualitatively similar time evolution is observed in experiments. Figure 6(c) shows a pore structure for $a=0.02,b_0=0,b_1=0.01,b_2=0.005,g=0.005$ and $V=2$ at $n=60000$.  By the effect of the surface tension, a fine structure does not appear and  the average interval between neighboring pores after the initial transient is larger than in the case of $g=0$. Furthermore, the growth velocity and the depth of the pore is smaller than in the case of Fig.~6(b). 
Figure 6(d) shows a pore structure for $a=0.045,b_0=0.0002,b_1=0.01,b_2=0,g=0$ and $V=3$ at $n=17000$. The top of the second interface dissolves and sinks downwards in time because $b_0$ is not zero.   

We can perform numerical simulation of a coupled map lattice model corresponding to the original model expressed by Eqs.~(2), (3) and (4). For this model, Eq.~(20) is replaced by  
\begin{equation}
q_{n+1}(i,j)=q_n(i,j)-[\alpha_0\exp\{k_dE_{en}(i,j)\}-\beta_0\exp\{k_oE_{en}(i,j)\}].
\end{equation} 
Figure 7(a) shows a pore structure for $a=0.02,\alpha_0=0.01,k_d=0,\beta_0=0.01,k_o=1$ and $V=2$ at $n=40000$. The Taylor expansion of Eq.~(21) by $E_{e}$ yields $b_0=0,b_1=0.01$ and $b_2=0.005$ in Eq.~(20), which corresponds to Fig.~6(b). Figure 7(b) shows a pore structure for $a=0.045,\alpha_0=0.05,k_d=0.2,\beta_0=0.0498,k_o=0$ and $V=3$ at $n=11000$. The Taylor expansion of Eq.~(21) by $E_{e}$ yields $b_0=0.0002,b_1=0.01$ in Eq.~(20), which corresponds to Fig.~6(d). Similar pore structures appear in the model Eq.~(21). Figure 7(c) shows a pore structure for $a=0.045,\alpha_0=0.01,k_d=1.2,\beta_0=0.005,k_o=1.1$ and $V=3$ at $n=10000$. The alumina film becomes even thinner by dissolution, and cusp-like structures rather than pore structures appear in the second interface.  
\section{Summary}
We have studied a simple model for spontaneous pore formation in anodic oxidation. It is one of typical problems for growing interfaces, however, this system is unique in that the interaction between two interfaces is important. The second interface between the oxide and the electrolyte is essentially unstable, because the interface moves toward the region of large electric field, and the first interface between the metal and the oxide is essentially stable, because the interface moves away from the region of large electric field. The interaction between the essentially stable first interface and the essentially unstable second interface makes a new kind of growth pattern. We have proposed a coupled map lattice model to understand qualitatively the time evolution of the anodic oxidation and performed several numerical simulations. We have reproduced some qualitative features of the anodic oxidation, such as the single pore solution, the dependence of the thickness $d$ on $V$, and the strongly nonlinear evolution of the pore structure. 

However, our model studied in this paper is a rather simplified one. We need to improve the model to incorporate the better boundary conditions using the  Butler-Volmer relation, the reaction-diffusion dynamics of some chemicals such as H$^{+}$ and OH$^{-}$. 
Especially, we would like to take an elastic effect into the coupled model to reproduce the regular array of pores.  We hope that a regular hexgonal array of pores might be obtained in the three-dimensional simulations of the generalized coupled map lattice model.       


\begin{thebibliography}{99}
\bibitem{rf:1} F.~Keller, M.~S.~Hunter, and D.~L.~Robinson, J. Electrochem. Soc. {\bf 100}, 411 (1953).
\bibitem{rf:2} H.~Masuda and K.~Fukuda, Science {\bf 268}, 1466 (1995).
\bibitem{rf:3} A.~P.~Li, F.~M\"uller, A.~Birner, K.~Nielsch, and U.~G\"osele, J. Appl. Phys. {\bf 84}, 6023 (1998).
\bibitem{rf:4} V.~Zwilling, E.~Darque-Ceretti, A.~Boutry-Forveille, D.~David, M.~Y.~Perrin, and M.~Aucuturier, Surf. Interface Anal. {\bf 27}, 627 (1999).
\bibitem{rf:5} J.~P.~O'Sullivan and G.~C.~Wood, Proc. Roy. Soc. Lond. A {\bf 317}, 511 (1970). 
\bibitem{rf:6} V.~P.~Parkhutik and V.~I.~Shershulsky, J. Phys. D {\bf 25}, 1258 (1992).
\bibitem{rf:7} S.~K.~Thamida and H.-C.~Chang, Chaos {\bf 12}, 240 (2002).
\bibitem{rf:8} G.~K.~Singh, A.~A.~Golovin, I~S.~Aranson, and V.~M.~Vinokur, Europhys. Lett. {\bf 70}, 836 (2005).
\bibitem{rf:9} G.~K.~Singh, A.~A.~Golovin and I.~S.~Aranson, Phys. Rev. B {\bf 73}, 205422 (2006).
\bibitem{rf:10} C.~Sample and A.~A.~Golovin, Phys. Rev. E {\bf 74}, 041606 (2006).
\bibitem{rf:11} {\it Dynamics of Curved Fronts}, ed. by P.~Pelc\'e (Academic Press, Boston, 1988).
\bibitem{rf:12} T.~A.~Witten and L.~M.~Sander, Phys. Rev. Lett. {\bf 47}, 1400 (1981).
\bibitem{rf:13} A.~Karma and W-J~Rappel, Phys. Rev. E {\bf 57}, 4323 (1998).
\bibitem{rf:14} H.~Sakaguchi and M.~Ohtaki, Physica A {\bf 272}, 15 (1999).
\bibitem{rf:15} H.~Sakaguchi, T.~Yoshida, S.~Nakanishi, K.~Fukami, and Y.~Nakato, J. Phys. Soc. Jpn. {\bf 75}, 114002 (2006).
\bibitem{rf:16} H.~Sakaguchi and R.~Baba, Phys. Rev. E {\bf 76}, 011501 (2007).
\bibitem{rf:17} P.~G.~Saffman and G.~Taylor, Proc. Roy. Soc. Lond. A {\bf 245}, 312 (1958).
\bibitem{rf:18} H.~Sakaguchi and K.~Noto, J. Phys. Soc. Jpn. {\bf 78}, 024601 (2009).
\end{thebibliography}
\end{document}